National Information Infrastructure Development in Canada and the U.S.: Redefining Universal Access

and Universal Service in the Age of Techno-Economic Convergence


Martin Dowding
Information Policy Research Program, University of Toronto
School of Library, Archival, and Information Studies
University of British Columbia, Vancouver, B.C.
<dowdingm@interchange.ubc.ca>


<div align="center">Abstract</div>


This exploratory and descriptive research compares the policy-making processes and policy

recommendations regarding universal service and universal access developed by the U.S. National

Information Infrastructure Advisory Council (NIIAC) and the Canadian Information Highway Advisory

Council (IHAC) in conjunction with related federal government agencies. Created in 1993 and 1994,

respectively, the Councils were charged with "bringing forward" the concepts of universal service and

universal access to adjust to the effects of deregulation, new and converged Information and

Communication Technologies (ICTs), and neo-liberal economic competition and globalization that

included acknowledging the private sector as the primary creator of the Information Highway. Both

Councils were informed by the history and discourse pertaining to Universal Service and telephony; they

received submissions from the public, private, and NGO sectors and individuals to complete their

investigation; both Councils delivered policy recommendations to their respective governments, pointing

to traditional information and public access institutions such as libraries, schools, and hospitals as the

most feasible points to begin to guarantee access to the Information Highway. As its primary data this

study relied on the Councils' preparatory briefing documents and their final reports as well as interviews

with Council members, government agency, and NGO members. This qualitative study used as its

methodology organizational, policy, narrative, and discourse analyses to create a picture of what

universal service and universal access were and what they became in the hands of NIIAC, IHAC, and

associated government agencies. The U.S. had started with a more clearly defined universal service

tradition than Canada, and undertook a more complex policy-making process with more experienced

personnel. It was also clear that IHAC had in many ways followed the U.S. model and arrived at many




similar recommendations as NIIAC. Because of the inevitability of technical, economic, and social change related to the Information Highway, no definitive outcome to the Universal Service and Universal Access "story" can be determined. Because the Canadian government did not follow up on some of IHAC's most crucial recommendations, the Canadian Information Highway "story," in particular, has been left less complete. It is concluded that because the Information Highway was to be primarily the responsibility of the private sector in a rapidly developing global economy, largely directed by the U.S., the Councils' recommendations would be inevitably similar. It was clear that the U.S. was setting a new "standard" for international ICT policy. It became clear, too, that because of de-stabilization introduced by convergence the meaning of "universality" itself had been seriously challenged. But it was also clear that, due to rapid and continuous technological, social, and policy change, the universal service and universal access definitions and discourse will continue to change and the discourse in which they are couched cannot come to a definitive close.

Introduction:

The conclusions reflected in this paper are drawn from my dissertation and research conducted for the Information Policy Research Program at the University of Toronto. The research compares policy-making processes and policy recommendations regarding universal service and universal access developed by the U.S. National Information Infrastructure Advisory Council (NIIAC) and the Canadian Information Highway Advisory Council (IHAC) and related federal government agencies. The Councils, created  in 1993 and 1994, respectively were charged with "bringing forward" the concepts of universal service and universal access to adjust to the effects of telecommunications deregulation, new and converged Information and Communication Technologies (ICTs), and neo-liberal economic competition and globalized communication. As NIIAC and IHAC began their deliberations considerable debate developed regarding the origins and purpose of Universal Service and Universal Access. For some advocates, for example librarians, Universal Service and Universal Access are the legacy of a tradition of social justice invoked to guarantee holistic social participation. The tenets of social welfare institutions such as public schools, libraries, health clinics, and community information centres, hold to that legacy,



having been founded on principles of universality. But for others, particularly the private sector, Universal Service and Universal Access were and remain a technical and economic (or techno-economic) means to create and distribute information for a profit. Universality is seen to have been foisted upon them by regulators and legislation that has come to be perceived as being out of step with globalized competition. If globalization continues to dominate our economic discourse, and if Universal Service and Universal Access are to be addressed at all, they must be negotiated through international economic agreements such as NAFTA, and groups such as the World Trade Organization. However, as many social critics concerned with issues of equity have pointed out, such economic treaties jeopardize national and regional legislation and regulation. According to the dominant economic discourse, equitable access can only be achieved through competition, not negotiated through social co-operation. At the behest of federal governments in most industrialized countries the Information Highway has been placed firmly into the hands of the private sector. In keeping with its ideological mandate, the private sector treats information and *especially* the *means* for transferring it, as commodities – useful for their exchange value rather than their use value. At the same time, the information and communication private sector, once subject largely to national regulation, has became responsible to an ever-more influential international market elite, and less responsible to national and regional interests. National information standards and policies are increasingly subject to the supra-national effects of agreements such as WTO, WIPO, or GATT. In other words, if globalization is becoming inevitable, so is comparative policy analysis.

According to Arnold Heidenheimer (1990, pp. 1-2) there are three primary reasons for conducting comparative policy analysis:

(1) To look for guidance in designing better policies.

Different countries will have different strategies for addressing similar problems. As many nations' communication strategies become more alike due to technical and international economic, regulatory, and policy convergence, our ways of coping will benefit from knowing what to do and perhaps, what not to do.



(2) A second reason for conducting comparative policy analysis is to understand better how government institutions and political processes operate as they deal with day-to-day problems. Regardless of how much change occurs in market and political dynamics, certain questions remain unchanged, such as who governs, how they govern, and who gets what once policy decisions have been made.

(3) A third reason for studying policies across national borders lies in the increasing importance of globalized interdependence. As information and communication and ICTs are increasingly relied on for international commerce, health care provision, education, and social interaction, access problems – and solutions – will become more alike and more common. To understand such problems and to come to common solutions will require comparable problem-solving techniques.

Geographically, this comparative policy study is confined to the United States and Canada as an important example of globalization, and as part of a long-standing political, cultural, and economic policy research tradition. Such reasons were articulated succinctly by IHAC in a report prepared shortly after a fact-finding mission to NIIAC headquarters in Washington in 1994. At that meeting the then head of the Information Infrastructure Task Force (IITF) Universal Service Working Group, remarked that "if Canada and America cannot figure out how to cooperate on global information highway issues then no two other countries can."(IHAC I, Visit Report, Oct. 6-7,1994, p. 4).[1] The implication here is that we are so much alike that we will both do the same thing to overcome the challenges posed by the Information Highway. On the other hand, other NIIAC documents suggest that the U.S. would do what they thought necessary to protect their own information interests, regardless of the approach of other nations.

Significance of the Research to Public Welfare and the Public Sector:

What, then, is the significance of this research to the non-profit sector? The rationale for my research is based on several important, fundamental issues pointed out by, among others, Karen Adams and Bill Birdsall in their book: *Understanding Telecommunications and Public Policy*
(1998, CLA): They refer to:



(1) Overwhelming influx of information and communication technologies (ICTs) in public sector institutions.

[Secondary but related: ICTs deliver materials owned by corporate vendors and not public sector institutions – which raises questions about ownership of information, privacy, security, and ethical use of information, to name some well-known access questions]

(2) The successful adoption of ICTs is contingent on having sustained access to telecommunications services.

[Budgets have had to be adjusted to accommodate sustainability costs – often jeopardizing or transforming public sector information policies and practices]

(3) Telecommunications globalization, deregulation, and technical convergence have created the need to redefine concepts such as universal access in an entirely new context

Research Questions:

I began my research with certain expectations and assumptions about how Canada and the U.S. would approach the issues at hand. IHAC's response, I thought, would be to highlight the differences between Canada and the U.S., by pointing to our tradition of social welfare, such as universality in healthcare and education, as justification for guaranteeing universality regarding access, even to the point of challenging the notion of competition. NIIAC, on the other hand, I assumed, would rely on its free-market tradition and promote competition even to the detriment of access [recently, for example, we have seen examples of injurious competition in electricity distribution in California].

My research questions were designed, in part, to seek out if and how disposed IHAC and NIIAC were to principles of just and equitable distribution – that is, how they would address universality. I saw that issue, especially, as an important bell-wether for the direction of National information infrastructure development in Canada and the U.S..

The research questions, then, are as follows:



1. What has been meant by universal service and universal access?

2. How and why have the original meanings been affected/challenged/de-stabilized by technological changes (such as convergence) and economic liberalization and globalization (such as free-trade) up to the point of the IHAC and NIIAC processes?

3. What did IHAC and NIIAC attempt to do to with regards to universal service and universal access? Why did they carry out their work the way they did?

4. How and how well were IHAC and NIIAC informed about universal service and universal access?

5. What were the outcomes/recommendations of IHAC and NIIAC regarding universal service and universal access?

6. What are the similarities and differences between the discourse of the two committees, in relation to universal service and universal access?

7. How can such similarities and differences be accounted for in relation to other national differences in information/communications policy and legislation?

Methodology:

My research, which is qualitative, relies on how the language, meaning, and narrative of information policy record and frame the transformation of a discourse. It is true that quantitative analyses for ICT use and implementation studies are valid for institutional use studies and as tools to determine the diffusion of technologies at local, national, and international levels. Today, for example, the NTIA in the U.S. and in Canada Industry Canada and Statistics Canada produce reports designed to assist in identifying the "digital divide." But such analysis, in the context of when this experimental policy was developed, is somewhat premature and inconclusive. Given the newness of the Information Highway, policy scholars and policy makers were at the stage of describing and exploring what might be possible and what ought to be done. And their exploration – at that stage – was defined and created through language and argumentation, in a variety of privileged and less-privileged discourses.

My study, then, is primarily an analysis of the textual narrative and discourse in Canada and the U.S. of the concepts of Universal Service and Universal Access. The study uses as its methodology



organizational, policy, narrative, and discourse analyses to create a picture of what universal service and universal access were and what they became in the hands of NIIAC, IHAC, and associated government agencies. The study explores how those meanings have been expressed in the two countries, in disparate discourses, and how they have transformed over time.[2] Central to the recent "access" and "service" discourse are the procedures and reports undertaken and prepared by the "Access and Social Impacts" sub-group of IHAC and the "Access to the NII" working group of NIIAC. Those sub-organizations and the documents they and their closely related organizational units[3] produced are some of the most conspicuous recent attempts to clarify policy requirements needed to redefine Universal Service and Universal Access as they relate to the information highway. They reflect policy negotiations and processes that are part of a long-term social, political, and economic discourse. But because of the "newness" of the converged information highway infrastructure, the inclusion of new policy makers, and the new kind of policy making required to negotiate the changed NII scenario, this study is an exploration and description of what policy analysts call an "explanatory policy foundation." To that end, this study is an exploration of how policy meaning is transformed and how policy-making is carried out within a specific organizational setting.

For my primary data I relied on the Councils' preparatory briefing documents and their final reports, much of which was archived on-line. As well, I carried out interviews with 18 Council members and NGO representatives. The materials collected amounted to more than 2,000 pages of preliminary documents, interim reports, public submissions, and the Councils' final reports which, in total, comprise approximately 500 pages. For my interviews, the transcripts of which amount to approximately 300 pages, I travelled to the cities where the Councils's home offices had been or to where many of the council members work, Washington, Toronto, and Ottawa. I telephoned those interviewees I could not visit personally.

Why Should Universal Service Change?

The relatively simple policy model of the regulated monopoly is now unworkable since what was relatively easily managed as Plain Old Telephone Service (POTS) under a monopoly must now be



managed in the form of Public Access to Network Services (PANS) in what is – at least called – a competitive market. The situation is made more complex, on one hand, by technical convergence (in multi-capacity ICT applications) and, on the other hand, by more diverse information transfer and communication through a variety of other new technologies such as cell phones, pagers, palm pilots, etc. Today, due to these factors, and the trend to globalization, and because perceived information "needs" have become more complex,[4] determining what should be accessible is extremely difficult, as is creating the policy required to determine how to guarantee accessibility.

NIIAC and IHAC: The Policy Makers:

It is significant that with the full encouragement of their respective federal governments, both Canadian and U.S. Councils acknowledged the private sector as the primary creator of the Information Highway. Both Councils were informed by the history – or the effect – of universal access to telephone service. Both NIIAC and IHAC received their respective federal government's mandates for approaching the issues at hand. Both received submissions from the public, private, and NGO sectors and individuals to complete their investigation; both Councils delivered policy recommendations to their respective governments, significantly pointing to traditional public access, non-profit institutions such as libraries, schools, and hospitals as the most feasible points to begin to guarantee access. However, unlike NIIAC which held open meetings across the U.S., it is significant that IHAC held its meetings, which were closed to the public, exclusively in Ottawa. Also significant was the fact that IHAC was exclusively responsible to only one federal government body, Industry Canada.

In the U.S., the origins and purpose of universal service have been more clearly articulated than in Canada – which is, as I have concluded in my research – the foundation for a stronger and more comprehensive U.S. analysis of the service and access situation. The U.S model derives more directly from telecommunications principles than Canada's model. The purpose of the adoption of universal service principles in the U.S. is interpreted in two ways and derives from two acknowledged traditions stemming from early telephony (1) as a kind of price of access to a profitable market – that is – an inconvenience undertaken to satisfy regulators; or (2) a socially based holistic effort to guarantee full



social participation – that is, to do more than simply connect users to an infrastructure.

The U.S. Telecommunications Act of 1996, significantly created during and after the NIIAC process, is the clearest articulation of how the U.S. proposes to address post-monopoly universal service issues. Legislators there were able to take advantage of the completed NIIAC deliberations before finalizing the new 128 page Telecommunications Act. Sections 254-255 of the 1996 Act comprise a precise, five-page portion committed exclusively to access issues. In those sections are clearly defined universal service principles and a complex series of definitions of the services to be guaranteed by the Act and how the Federal Communications Commission will address changes to universal service when it is necessary.

In Canada's Telecom and Broadcasting Legislation, both significantly developed some years before IHAC's research was started, access is considerably sketchy in comparison. The term "Universal Service" is not used at all. In the Telecom Act, which amounts to 35 pages, the Acts' "objectives" include a few references to guaranteeing service for the sake of sovereignty and affordability. The Canadian Broadcasting Act, only a page longer than the Telecom Act, guarantees Canadian ownership and access to as wide a variety of cultural and social issues as is feasible. But it provides very little detail on how that should happen. Having enshrined more thoroughly informed access issues in legislation has meant a more obviously conscientious commitment in the U.S.

Canada's policy model has to a large degree relied on corporate telecommunications principles as well. But the foundation for the Canadian rationale for universal service and electronic media is found most clearly in early broadcasting deliberations such as the Aird Commission (1929). The Aird report, which received more than 700 briefs in 26 public hearings held across the country, was largely influenced by Canadian nationalism. At that time, the U.S. was making private media the norm, while the forerunner to the Canadian Broadcasting Corporation was being developed to counteract the influence of U.S. culture carried over the airwaves. While Canada founded its communication policy on nationalism, based on the culture inherent in Canadian content, it now appears to have been drawn away from the broadcasting tradition in order to adopt the telecom and telephony-related U.S. model – which prompts questions such as, how different are we culturally? How different will we remain? How



has Canada's policy discourse changed? What is the effect of globalization?

Combined with other collaborating federal agencies, NIIAC and IHAC defined and redefined Universal Service and Universal Access with a primarily techno-economic discourse which I have framed within a narrative process. Those organizations and agencies looked to the historical legacy of universal service and universal access in preparation for their work, although as I have suggested, the relatively clearly stated tradition in the U.S. considerably influenced the Canadian process. I have analysed what they received (for the beginning of their deliberations), what they did during their deliberations (in the middle of the policy-making process), and what recommendations they made in their final reports (as the end of their narrative discourse). In this paper I concentrate on the "beginning" of the story and how the beginning, born out of a former social, economic, policy, and technical regime, remains so problematic as we attempt to fashion an "ending" for the access narrative.

Because there is a de facto diminished feasibility in universal access and universal service under post-monopoly conditions, much recent policy-making is, in effect, experimental. Policy development of this nature is exploratory, and speaking in terms of public policy theory, develops from the "statement" stage. According to Parsons (1995, pp. 85-6) this type of analysis:

> Is concerned with how problems are formed and framed, and how they become – or
> do not become – items or issues on the policy agenda.

To further explain this social or public policy concept Parsons also cites U.S. policy scholar Hugh Heclo (1974) when he says policy-making in this experimental sense may be viewed as a form of "collective puzzlement on society's behalf; it entails both deciding and knowing" (Heclo 1974, p. 305). One of the most important issues at stake in the IHAC and NIIAC process, however, was who made up the "collective" that was doing the "puzzling." Despite the inclusion of Council members devoted to access to information as a social justice issue (for example, teachers, librarians, and other public sector and NGO representatives), most represented corporate interests. Such an imbalance has led many observers of the process to assume that, despite IHAC and NIIAC recommendations to address



"market failure" with guarantees of accessibility, neo-liberal mechanisms were paramount and unresolvable. The dominant membership and influence of the private sector also prompted some council members to speak out.

Universal Access and Universal Service as Narrative Discourse:

Introduction:

In this section I speak about what the two Councils received as the groundwork for bringing forward the principles of Universal Access and Universal Service (as the "beginning" of the narrative discourse). I then speak briefly to the documentation developed throughout the IHAC and NIIAC deliberations (to determine how the discourse was transformed, as the "middle" of the narrative). I discuss the final reports prepared by IHAC and NIIAC (to determine the "end" of the narrative and to determine if a new discourse "threshold"[5] emerged from the process). I also  address the principle findings from interviews with members of IHAC, NIIAC, and other government and non-governmental agencies in Ottawa and the U.S.. I conducted the interviews to determine if the documentary evidence I relied on is contradicted, or has in retrospect been discarded; I also wanted to determine whether new meanings of Universal Service and Universal Access had entered the discourse.

Both Councils' mandates made it clear they were responsible for updating the concept of Universal Access and Universal Service as a means – in part – to address the neo-liberal economic concept of "market failure." Both ideas had been couched in the discourse of monopoly-based public policy, regulation, and legislation in one form or another since the introduction of the telephone. Unfortunately, the purpose and origins of "universality" in the monopoly market, which have been eclipsed by post-monopoly conditions, were what NIIAC and IHAC had to begin with. Breaking that mind-set and introducing a new one was a formidable task.

NIIAC: The beginning: What was received:



The U.S. tradition of defining Universal Service is almost as old as the technology – the telephone – that prompted the need for a definition. Two traditions emerged. One tradition was based on AT&T President Theodore Vail's 1909 defence of a monopoly, "One system... one policy... universal service," which Milton Mueller (1997) considers to be primarily a corporate, rather than a social justification for universal service. The other tradition derives, in part, from a brief statement in the U.S. communications Act of 1934, which has been interpreted as promoting social inclusion. For Mueller (1997), that second tradition is based on a myth perpetuated by mis-readings of Vail's statement, the 1934 legislation, and subsequent legislative and regulatory initiatives in the U.S.

Despite some doubts about the efficacy and interpretation of particularly the 1934 legislation (Mueller, 1997) the longevity of the traditions and the continuity of at least a working definition provided U.S. policy makers concerned with access issues with a foundation to revise, or to reject and attempt to build new definitions.

To prepare the U.S. Council members for their mandate, the National Technical Information Administration (NTIA) prepared *National Information Infrastructure Agenda for Action,* a document notable for its optimism. Then U.S. Secretary of Commerce Ron Brown noted that *Agenda for Action* was to be a "blueprint" for the NII. In doing so he was referring largely to a particular part of the document, "Principles and Goals for Government Action," which reports that the IITF was:

> undertaking a wide ranging examination of all issues relevant to the timely development and growth of the National Information Infrastructure. Specific principles and goals in areas where government action is warranted have already been identified and work has begun....

The first Principle and Goal cited in *Agenda for Action,* and *the* fundamental goal to which neo-liberal economies like the U.S. (and Canada) were aspiring, was to "Promote Private Sector Investment" in order to "expand competition in communications and information markets." Indeed, it is that principle and goal which leads, as a not very surprising corollary, to the second Principle and Goal cited in



*Agenda for Action*, the need to "Extend the 'Universal Service' Concept to Ensure that Information Resources are Available to All at Affordable Prices." The writers of the document recognized the potential for harm to be done to accessibility by increasing competition.

To undertake that task of extending universal service NIIAC set about defining national goals for Information Highway implementation and gathering information on what a wide variety of commercial, cultural, and social constituencies in the U.S. felt were the priorities in guaranteeing access to information services and technologies. To begin with, as indicated in early correspondence from NIIAC staff members (June 27, 1994), NIIAC's sub-group Access to the NII, developed a bibliography including research, reports, and other data on the subject of universal service. They then conducted an "information environmental scan" of available services and information provided commercially; they conducted an international environmental scan to see what other countries had done regarding universal service; and they began efforts to define and clarify the term "universal service."[6] Just as important was the parallel and very influential group, the Information Infrastructure Task Force (IITF) created by the U.S. Federal Government which entertained public presentations and submissions throughout the process, during meetings held in Washington and across the U.S.

Two ground-breaking reports to which NIIAC had access were *Notice of Inquiry on Universal Service and Open Access* (1994) and *NII Field Hearings on Universal Service and Open Access: America Speaks Out* (Sept. 1994). Those documents emerged from efforts undertaken in 1993-94 by IITF's Universal Service Working Group of the Telecommunications Policy Committee. Inspired by direction in *Agenda for Action* NTIA directed the IITF to make a "special effort" to hear from public interest groups. A series of public hearings on universal service/open access and the NII was held and IITF heard from more than 230 witnesses, attracted more than 1,000 attendees, and produced more than 1,400 pages of testimony. The objectives of the hearings were twofold:

(1) to gather information on the best characteristics of an expanded concept of universal service; and (2) to provide important input to the Telecommunications Policy Committee of the IITF, working in conjunction with NIIAC and state regulatory commissions, to determine how the concept should be applied in the 21st century.



(NTIA, 1994-29, p. 4)

The hearings were frequently accompanied by technology demonstrations that both illustrated technical advances and policy issues relating to access. As it is reported in *America Speaks Out*, hearing locations were chosen by NTIA to reflect access needs and opportunities in different socio-economic areas of the country. Albuquerque, New Mexico was chosen for its:

> largely rural nature, wide-ranging demographics, and relatively low [telephone] penetration rates. South Los Angeles proved attractive as a forum for discussing universal service in an urban, inner-city setting with diverse ethnicities.... Durham [N.C.] was chosen for its 'high-tech' economy with largely rural environs.... 'Silicon Valley' (Sunnyvale) as representative of a technology-based economy in more urban surroundings... In Indianapolis NTIA sought to focus discussion on both universal service and open access ... where the economy is being... transformed from its traditional heavy industry and agrarian underpinnings to one heavily reliant on information and telecommunications. (NTIA, 94-29, p. 5).

From the hearings were developed the basic themes and core issues presented to NIIAC when it began its deliberations. The rationales for choosing the site locations also reflect many of the access concerns central to public welfare institutions which rely on access to affordable telecommunications – issues that have persisted in the new Information Society and Information Economy, despite and *because* of political and technological change.

IHAC: The Starting Point in Canada

What, then, was IHAC's first impression of Universal Access and Universal Service and how did it come to it? As I just pointed out, NIIAC received its *Agenda for Action*. IHAC received *The Canadian Information Highway: Building Canada's Information and Communications Infrastructure* (April, 1994) from Industry Canada.



*Building Canada's Information and Communications Infrastructure* embodies the fundamental vision and strategy the Canadian Government foresaw for the Information Highway at the beginning of the IHAC planning process. Cited in IHAC's first major report, *Connection, Community, Content* (1995) as the origin of the issues IHAC was to work with, *Building Canada's Information and Communications Infrastructure* includes Industry Canada's "National Vision and Strategy" and a brief synopsis of Canada's communication history, in some ways echoing the NIIAC reference to the U.S.1934 Communications Act, but more descriptive of the technical infrastructure than the policy and legislation that directed its creation:

> Since the invention of the telephone by Alexander Graham Bell in 1876, Canada has always considered it essential to have strong communications networks – owned and controlled by Canadians -- for trade, cultural vitality and national sovereignty. The government has actively supported the development of such networks: The Trans-Canada Telephone System was set up in 1932, the Canadian Broadcasting Corporation (CBC) was established in 1936; the transcontinental microwave networks were built in the late 1950s; cellular telephone services were rolled out in the late 1980s; and fibre optic networks have been laid coast to coast over that last two years. (Industry Canada, April 1994, p.7)

Corresponding policies, regulation, and legislation have been behind those efforts, including the Telecommunications Act (1993), the Broadcasting Act (1991) and the work of the Industry Canada and the CRTC, but significantly in a less concerted effort than in the U.S..

One of the early discussion papers produced by IHAC itself was "Access and Social Impacts Working Group (June 15, 1994): Working Group Discussion on Public Consultation." That paper states the Working Group's objectives regarding access. It observes that issues such as universal access, affordability , equity, and just and affordable rates are not new to communications. The discussion paper makes 30 recommendations, presented in five sets, about what to do today, in a



period of technical, policy, and regulatory change.

> One set is concerned with public policies to address universal and requisite services.... a second set is concerned with access and affordability with respect to telecommunications-based services.... a third set deals with access and affordability with respect to content-based services.... another set of recommendations addresses special measures for individuals... a final set of recommendations deals with special measures for remote and rural locations." (Access and Social Impacts Working Group, June 15, 1994, p. 5).

Recommendations: IHAC and NIIAC

In the U.S. NIIAC's recommendations regarding Universal Service are published in *Common Ground: Fundamental Principles for the National Information Infrastructure* (March, 1995) and *A Nation of Opportunity: Realizing the Promise of the Information Superhighway* (January, 1996). Central to the NIIAC recommendations were developing technical mechanisms for affordability and a time-frame for connecting all schools and libraries by the year 2000 and every individual by the year 2005. Affordability was addressed, as I have pointed out, in the Telecommunications Act, while universal connectivity remains somewhat elusive. It is significant, however, that in the U.S. the NTIA and the Communications NGO, the Benton Foundation, at the behest of the government, are monitoring the situation much more closely than Canada. From time to time the NTIA produces reports entitled *Falling through the Net* which provide primarily statistical analyses of which socio-economic and ethnic groups are, or are not information "have nots." The Benton Foundation policy analyses and reports on government decisions appear on their web site.

IHAC's recommendations for guaranteeing Universal Access, which was its preferred term, are found in its two final reports, *Connection, Community, Content: The Challenge of the Information Highway* (September, 1995) and *Preparing Canada for a Digital World* (September, 1997). There



you will find many recommendations similar to NIIAC's, including a request that the government follow up on IHAC's recommendations, in particular that the Canadian Government articulate a National Access Strategy by the end of 1997 and that it create a body to monitor the application of IHAC's recommendations. Neither of those recommendations was ever implemented by the Federal Government. Statistics Canada produces statistical reports on ICT use by community, age and income group, but it not as fine-grained as the NTIA. And as we do not have as developed an NGO tradition in Canada, it unlikely that we can have as clear an analysis of who has what access in Canada.

Results/Conclusions

Contrary to my expectations, IHAC did not carry out the policy process with the primary intention of universal access to information as a social justice issue – as Canada has traditionally done so, for example, for the sake of universal healthcare. My research shows that there are several reasons for coming to that conclusion. They include:

· NIIAC's relationship with other U.S. agencies, such as NTIA and IITF was much stronger than IHAC's relationship with other government agencies and they were, therefore, better informed about the issues.

· Because the Information Highway was to be primarily the responsibility of the private sector in a rapidly developing global economy, the Councils' recommendations would be inevitably similar, and therefore to the advantage to the U.S., whose economy Canada relies on for its stability.

· It was clear that the meanings of universal service had been more markedly transformed – or been more obviously clarified – in Canada because it had further to go in the first place. It was also clear that the U.S. discourse was setting the "standard" by which Canada, at least, was developing Information Highway policy.



· The U.S. has had a more clearly defined universal service tradition than Canada, and undertook a more complex, broad-based policy-making process with more experienced personnel – but came to a new "service" definition, rather than an "access" definition with the purpose of guaranteeing full social participation.

· IHAC had followed the U.S. model and arrived at many similar recommendations as NIIAC, and in doing so had abandoned traditional social welfare traditions. But it was also clear that, having followed only some of the U.S. example, Canada was destined to be unable to fulfill many of IHAC's recommendations.

In the minds of many critics of the IHAC process the most significant of Canada's failings after the IHAC recommendations has been the failure by the Federal Government to create a National Access Strategy and a national committee to oversee such a strategy.

For the final part of my dissertation research I conducted interviews with NIIAC, IHAC members, federal government agencies, and NGOs responsible for contributing submissions to the deliberations. Strikingly, everyone interviewed said the same thing about competition and monopolies and, despite there being ideological differences:

competition is far superior and has proven to be a good thing - great innovation has emerged. The Information Highway should remain as unfettered as possible.

One of the persons I interviewed was a member of the NIIAC sub-group "Access to the NII." He introduced himself as a liberal left democrat committed to social justice. But he also expressed great joy at having been responsible for putting a stop to any taxation of the Internet, citing how dangerous a precedent such a move would be. He said that because there are now many more ways of gaining access to information, surely access is now easier. But, at the same time, the recommendations the



policy bodies developed, particularly IHAC – as they were delivered in the final reports – couldn't haven't solved the issue about which I am most concerned – sustainability of access – which happens to be the object of my ongoing and future information policy research.

It is important to remember that "universality" was much more feasible under monopoly conditions, prior to technical convergence. Someone – or some entity, namely the Telecommunications companies – historically agreed to accommodate access as a social justice concern, despite their primary objective of profit. Paradoxically -- with the move to supposedly superior information technologies in supposedly superior globalized markets – there is less assurance of universal access than there was under the much maligned monopolies. Outside of the marketplace, access takes on an urgency caused by the problem of sustainability. Access considered in these terms might refer to sustained guaranteed access to personal or institutional information transfer methods, procedures, and technologies

In Canada, that concern prompted a considerably negative response from some observers and participants in the process.

IHAC member and union leader Jean-Claude Parrot refused to sign IHAC's final report, and submitted his own minority report. In my interviews with members of IHAC and NIIAC I heard many anecdotes about the army of lawyers that accompanied private sector council members at meetings while librarians, teachers, and academics arrived on their own or, at best, with a student assistant.

Implications for Further Research:

As I pointed out earlier, the public policy-making undertaken by IHAC and NIIAC was to some degree experimental. Perhaps it should be viewed as general, tentative, or incomplete. But as a policy exercise it has, at the very least, brought us to a new threshold in the access discourse, and a stepping-off point for further research. Information and communication initiatives in Canada and the U.S., developed since the IHAC and NIIAC process, are still subject to questions such as who governs, how they govern, and who gets what once policy decisions have been made. The private sector,



governments, and NGO groups continue the universal service and universal access debate to address changing information needs, new policy requirements and constantly changing technology in the context of the Information Economy, or as I prefer to call it -- the Information Society. But the discourse has been transformed, in my estimation, to favour the U.S. corporate, technical, and content convergence.

A Contemporary Addendum:

My study does not address the situation of universal access and universal service after the NIIAC and IHAC final reports were written, except to point to obvious complications, or failures to do what was evidently deemed necessary in the NIIAC and IHAC's deliberations and recommendations. However, in closing, as an after-word meant to justify further research in this area, and to provide further evidence that NIIAC and IHAC's recommendations regarding universal service and universal access and the private sector were unworkable, I will add the following observations.

Today, the private sector, called upon by NIIAC and IHAC to build the Information Highway, claims the cost is too high to do so without more government assistance. E-Rates in the U.S. have become problematic, primarily because of the role of the private sector in "deciding" what service is "worthy" of support.[7] The media and telecommunications corporations (the ultimate recipients of the funds), in effect, act as "middle-men" and maintain considerable influence in how the money they receive is spent. The private sector has, in effect, no more *incentive* today to guarantee universality than they did in the time of monopoly-State regulation. They may, in fact have less incentive, as there is a tendency on the part of the State to abstain from regulation in the face of open competition.

What has been adopted in Canada is also proving to be problematic. Recently, private sector interests responded negatively to a Federal Canadian report on the cost of developing the infrastructure required to bring to fruition IHAC's promises that by 2004 a system would be in place to allow Internet access to all Canadians, in urban areas as well as rural areas. The Broadband Task Force, a techno-economically based policy group struck in 2000 was, significantly, headed by the former head of IHAC, David Johnston; the dominant membership was senior executives of telecommunications companies.



When the Broadband Task Force report was released in June, 2001 responses from telecommunications industry and Internet suppliers, the very groups said to be responsible for the creation of the Information Highway, voiced their concern that there would not be enough competition to fulfill the report's aspirations. Comparing the Internet to Canada's first communication myth inspired by the cross-country railroads, something we Canadians call the "National Dream," an Ottawa journalist summed up the situation in the following way, after the Task Force report was released:

> Ironically, in addition to linking Canada from coast to coast, government funding of the
> national railway in the late 1880s also led to a financial scandal for the country's first
> federal government. And some critics of public investment in network access today say
> they worry that a handful of incumbent telecommunications companies could siphon the
> lion's share of the funds (Bonisteel, June 18, 2001).



For Canada, then, there is a new but familiar discourse which is, like the origins of the railway suggest, not exclusively but fundamentally techno-economic. However, the narrative is far from complete. My research argues that there is no true ending to the universal access and universal service "story," either as a techno-economic issue or a socio-technical issue. There are, instead, two incomplete or non-stories:

    (1) universal service and universal access in the context of techno-economic policy-making, confounded by technical change and market uncertainty, and

    (2) universal service and universal access in the context of socio-technical policy-making, confounded by technical change and market uncertainty.

There is, then, no ending to the opportunity and obligation to survey and evaluate the issues, meanings, and effects of universal service and universal service in relation to the social, technical, and economic change they represent.

---

Notes

---

1. Irving also pointed out the long, "robust tradition" of bilateral cooperation between the two countries but at the same time citing differences in culture, perhaps the greatest sticking point when we ask to what are we guaranteeing access?.

2. As this thesis is concerned, on the most part, with electronically transferred information, the proposed time frame for the study coincides, roughly, with the period of the 20th century. I will close my time frame at the end of the 20th century, as it represents a symbolic "closure" as well as the period in which the first significant responses to IHAC and NIIAC were becoming evident. Those responses, particularly the U.S. follow-up on "E-Rates" and the "Connecting Canadians" initiative, will be considered the "end" of the 20th century narrative structure of the creation of the Information highway.



3.NIIAC, IITF and the NTIA, and IHAC, Industry Canada.

4.The question of more complex "needs" arises because, as new ICTs are adopted and adapted they sometimes eclipse traditional means of access, providing new and exclusive means and methods for social procedure such as communicating with government, or simply locating a product or service previously locatable with a telephone or in hard-copy format at a well-known institution.

5.I speak of a "threshold" in reference to the possibility of definitive change in the discourse.

6.At this point the term "universal information infrastructure access" was put forward but was not used again.

7.Perhaps even more problematic is that, in the U.S., public funding for Information Highway access in schools and libraries is frequently contingent on those institutions agreeing to adopt Internet filtering devices, diminishing access on ideological grounds and not promoting universality.